\documentstyle[aps,multicol,epsf]{revtex}

\begin{document}


\draft

\title{\textit{Ab initio} quasiharmonic equations of state \\ for
dynamically-stabilized soft-mode materials}

\author{N.~D.~Drummond and G.~J.~Ackland}

\address{Department of Physics and Astronomy, The University of
Edinburgh, JCMB, The King's Buildings, Edinburgh, UK}

\date{\today}

\maketitle

\begin{abstract}

We introduce a method for treating soft modes within the analytical
framework of the quasiharmonic equation of state.  The corresponding
double-well energy-displacement relation is fitted to a functional
form that is harmonic in both the low- and high-energy limits.  Using
density-functional calculations and statistical physics, we apply the
quasiharmonic methodology to solid periclase (MgO). We predict the existence
of a B1--B2 phase transition at high pressures and temperatures.

\end{abstract}

\pacs{61.50.Ks, 64.70.Kb, 71.20.-b}

\begin{multicols}{2}

\section{Introduction}

The quasiharmonic approximation \cite{Wallace} provides a means of
extracting finite-temperature properties of materials from static
calculations. It assumes the vibrational properties can be understood
in terms of excitations of non-interacting harmonic normal modes: {\it
phonons}. Lattice dynamics \cite{born} can be used to calculate phonon
energies by evaluating the eigenvalues of the dynamical matrix, which
involves second derivatives of the crystal energy with respect to
atomic displacements.

The frequencies of these modes depend on the crystal's density. Hence
they have a temperature-dependence that arises simply because of
thermal expansion in the material.

Recent developments in {\it ab initio\/} energy calculations have
enabled full phonon dispersion curves to be obtained, leading to a
resurgence of interest in the quasiharmonic approach. Difficulties
arise, however, when the dynamical matrix has negative eigenvalues,
indicating that the crystallographic structure is not a local minimum
of energy.  Such crystals may be {\it dynamically stabilized}: because
of their large entropy, they may represent a minimum of free energy at
high temperature. Within the conventional assumption of harmonic
phonons, the quasiharmonic framework leads to divergent free energies
in these cases. Such systems have been treated numerically from first
principles using Monte Carlo methods with effective Hamiltonians
\cite{effective} or molecular dynamics simulations of a reduced set of
modes \cite{warren_minmag}.  Here, we relax the quasiharmonic
assumption of harmonic modes while retaining the approximation of
non-interacting phonons and show how the intrinsic anharmonicity of
such modes can be included in analytic free energy calculations.

The mineral {\it periclase\/} (MgO) is of some geological significance
as one of the supposed constituents of the Earth's lower mantle. It is
generally believed that along the geotherm---the conditions of
pressure and temperature actually occurring in the mantle---periclase
remains in a single phase. Under other conditions, however, previous
calculations have suggested that periclase has two phases in its solid
state: a sodium chloride-like face-centered cubic phase (B1) and a
cesium chloride-like simple cubic phase (B2) that is favored at
extremely high pressures \cite{karki}.

Imaginary phonon frequencies are found in the B2 phase of periclase
and have attracted enormous attention in the other principal
constituent of the lower mantle, magnesium silicate perovskite
(MgSiO$_3$)
\cite{warren_pcm,warren_phd,MgSiO3_DAC,parlinski_mgsio3,aro:perov1}.

Equilibrium structures, thermodynamic properties and compositions
depend on the free energy. Here we present a calculation of the free
energy of periclase as a function of density and temperature. We use
the pseudopotential-plane wave approach to evaluate total energies and
the method of finite displacements to evaluate pressure-dependent
force constants \cite{ackland}, including effective charges and
dielectric constants \cite{parlinski_mgo} for the longitudinal optic
modes. Based on calculated phonon frequencies and the quasiharmonic
approximation, we present a first-principles calculation of the phase
diagram and thermodynamic equation of state of solid periclase: the
relationship between pressure, density and temperature.

\section{The quasiharmonic method}

\subsection{\textit{Ab initio} calculation of specific Helmholtz free energies}

The first stage of the calculation is to obtain the specific (with
respect to mass) Helmholtz free energy of each phase as a function of
density and temperature. We write the free energy as
the sum of the frozen-ion interaction energy and the free
energy due to lattice vibrations.

\subsubsection{The frozen-ion energy \label{section_spec_energy}}

The frozen-ion energy---the interaction energy of the crystal with the
ions fixed in their equilibrium positions---is, by definition,
temperature-independent. Hence the free energy is simply equal to the
internal energy. In order to determine the dependence on density,
total energy densiyty functional calculations are carried out for each phase at a range of
different lattice parameters.

\subsubsection{The lattice thermal Helmholtz free energy
\label{section_latt_thermal}}

We calculate the lattice thermal Helmholtz free energy of each phase
as a function of density and temperature within the framework of the
harmonic approximation \cite{ashcroft}.

For a range of lattice parameters, we evaluate the matrix of force
constants for a supercell (several unit cells) of the phase under
consideration, subject to periodic boundary conditions. We evaluate
the forces on the ions in a crystal when one ion is displaced slightly
from its equilibrium position: from such calculations the matrix of
force constants may be constructed\cite{ackland}.

We denote the matrix of force constants by $\phi$, where
$\phi_{l,n,\alpha;m,p,\beta}$ is the component of force in direction
$\alpha$ on ion $n$ in unit cell $l$ when ion $p$ in unit cell $m$ is
displaced infinitesimally in direction $\beta$, divided by the
magnitude of the displacement.

Pairs of density-functional calculations are carried out with an ion
displaced from equilibrium along one of the Cartesian axes by a small
amount in first a positive and then a negative sense. By averaging the
resulting Hellmann-Feynman forces on the ions from the first
simulation with the negative of the forces from the second,
first-order anharmonic contributions to the force constants are
eliminated.

The set of rotations under which the crystal structure is invariant
are identified and the rotation matrices, together with the mappings
between the ions under the symmetry operations, are evaluated. For a
given pair of ions $(l,n)$ and $(m,p)$, the matrix of force constants
$\phi_{l,n,\alpha;m,p,\beta}$ transforms as a second-rank tensor.
However, for a symmetry operation, the transformed matrix must be the
same as the (unrotated) matrix of force constants between the pair of
ions which are mapped to $(l,n)$ and $(m,p)$. Hence, new elements of
the matrix of force constants can be obtained by application of these
point symmetries. Translational symmetries can be identified and
exploited in a similar fashion.

The matrix of force constants should be symmetric \cite{ashcroft} and
Newton's third law must be satisfied: if an ion is displaced slightly
then the restoring force on that ion must be equal and opposite to the
total force on all of the other ions. Hence we must have:

\begin{eqnarray} \phi_{l,n,\alpha;m,p,\beta} & = &
\phi_{m,p,\beta;l,n,\alpha} \\ \phi_{l,n,\alpha;l,n,\beta} & = &
-\sum_{(m,p) \neq (l,n)} \phi_{l,n,\alpha;m,p,\beta}. \label{eq:phicon}
\end{eqnarray}

The force constants are obtained from separate numerical calculations;
hence small violations of these requirements may occur. These two
conditions are therefore alternately imposed on the matrix of force
constants until further application leaves the matrix unchanged
\cite{ackland}.

The next stage of the calculation involves the construction of
dynamical matrices \cite{ashcroft} for various wavevectors in the
first Brillouin zone. Diagonalization of the dynamical matrix for a
given wavevector gives the spectrum of corresponding
eigenfrequencies. Strictly, these are only exact when the wavelengths
are commensurate with the dimensions of the supercell
\cite{ackland}. However, provided that the resulting dispersion curves
are smooth, it may be assumed that the interpolation errors are
negligible.

Cochran and Cowley \cite{cochran} have shown that the elements of the
dynamical matrix for an ionic crystal can be written as the sum of a
term that behaves analytically as the wavevector tends to zero and a
term that is non-analytic at the zone center. The latter term vanishes
as the boundary of the Brillouin zone is approached. This term arises
because longitudinal optic (LO) phonons cause an electric polarization
field to be set up within the crystal as the oppositely-charged ions
are displaced in opposite directions. The resulting long-range
interactions cannot be calculated within the framework we have
described so far because of the limited size of the simulation
supercell. At the zone center itself, the LO phonon sets up a uniform
electric polarization that is incompatible with the periodic boundary
conditions on the supercell.

Cochran and Cowley's expression for the dynamical matrix is:

\begin{eqnarray} \tilde{\phi}_{n,\alpha;p,\beta}({\bf k}) = & &
\tilde{\phi}^N_{n,\alpha;p,\beta} ({\bf k})+ \frac{4 \pi
e^2}{\Omega|{\bf k}|^2 \sqrt{M_n M_p}} \nonumber \\ & \times &
\left(\sum_{\gamma=1}^3 k_\gamma Z_{n,\gamma,\alpha}({\bf k})
\right)^\ast\epsilon_0^{-1}({\bf k}) \nonumber \\ & \times & \left(
\sum_{\gamma=1}^3k_\gamma Z_{p,\gamma,\beta} ({\bf k}) \right),
\label{equation_coch_cow}
\end{eqnarray}

\noindent where ${\bf k}$ is the wavevector, $e$ is the electronic
charge, $\Omega$ is the volume of the unit cell, $M_n$ is the mass of
ion $n$, $Z_{n,\alpha,\beta}({\bf k})$ is the Born effective charge
tensor for ion $n$ and $\epsilon_0 ({\bf k})$ is the electronic
(frequency dependent) dielectric function. The first term on the
right-hand side is the component of the dynamical matrix that is
analytic as ${\bf k} \to 0$, while the second term is the non-analytic
part due to macroscopic polarization effects. We use our matrix of
force constants evaluated using the Hellmann-Feynman theorem in a
cubic supercell to evaluate the analytic part as:

\begin{equation} \tilde{\phi}^N_{n,\alpha;p,\beta} ({\bf k}) = \frac{1}{\sqrt{M_n
M_p}} \sum_m \phi_{0,n,\alpha;m,p,\beta} e^{-i {\bf k} \cdot ({\bf
R}_0 -{\bf R}_m)}, \end{equation}

\noindent where ${\bf R}_m$ is the position vector of unit cell $m$.

Following Parlinski {\it et al\/}\cite{parlinski_mgo} we assume that the second term on
the right-hand side of Equation \ref{equation_coch_cow} falls off from
its value at the Brillouin zone center with a Gaussian profile. For
wavevectors in the first Brillouin zone this term is:

\begin{eqnarray} & & \frac{4 \pi e^2}{\Omega |{\bf k}|^2 \sqrt{M_n
M_p}\epsilon_0({\bf 0})} \left( \sum_{\gamma=1}^3 k_\gamma
Z_{n,\gamma,\alpha} ({\bf 0}) \right)^\ast \nonumber \\ & \times &
\left(\sum_{\gamma=1}^3 k_\gamma Z_{p,\gamma,\beta} ({\bf 0}) \right)
\times e^{-\left(\frac{| {\bf k} |}{\rho_0 \kappa^{1/2}} \right) ^2 },
\label{eq:non_analytic} \end{eqnarray}

\noindent where $\kappa^{1/2}$ is the distance from the center to the
boundary of the Brillouin zone along the $k_x$-, $k_y$- and
$k_z$-directions and $\rho_0$ is a parameter determining the rate at
which the term falls off as the edge of the Brillouin zone is
approached.  Following Parlinski, we set $\rho_0 \equiv 1.2$.

The frequency density-of-states function is evaluated using the method
of Swift \cite{swift_phd} in which the Brillouin zone is sampled using
Monte Carlo methods. For a single harmonic mode of frequency $\omega$,
the Helmholtz free energy is given by:

\begin{equation} F_1(\omega) = k_B T \log \left( e^{\beta \hbar \omega
/2} - e^{-\beta \hbar \omega / 2} \right),
\label{equation_harmonic_free_energy} \end{equation}

\noindent where $\hbar$ is the Dirac constant, $k_B$ is Boltzmann's
constant, $T$ is the temperature and $\beta = 1/k_BT$. Hence, by
numerically integrating the product of the specific density-of-states
with the mean free energy of a normal mode, the specific lattice
thermal free energy can be calculated for a range of temperatures. The
dependence on density is found by interpolating between the results at
different lattice parameters.

\subsection{Polymorphism \label{section_polymorphism}}

\subsection{The Gibbs free energy}

We have calculated the Helmholtz free energy $f(v,T)$ as a function of
temperature $T$ and specific volume $v$ (the reciprocal of the
density). However, the appropriate thermodynamic potential for
constructing the $(p,T)$-phase diagram and evaluating the polymorphic
equation of state is the specific Gibbs free energy $g(p,T)$, where
$p$ is the pressure. The Gibbs free energy function for each phase can
be evaluated using the Legendre transformation:

\begin{equation} g(p,T)=f+pv=f-\left( \frac{\partial f}{\partial v} \right) _T v. \end{equation}

\subsection{The phase diagram}

Under conditions of fixed pressure and temperature, the system
consists entirely of the available phase with the lowest Gibbs free
energy. Thus the phase diagram in $(p,T)$-space can be evaluated.

\subsection{Combining phases}

For each pressure and temperature, we may evaluate the polymorphic
Gibbs free energy $g_{\rm poly}(p,T)$ as the lowest of the Gibbs free
energies for each phase. Given this, we may carry out a Legendre
transformation to the polymorphic Helmholtz free energy:

\begin{equation} f_{\rm poly}(v,T)=g_{\rm poly}-pv=g_{\rm poly}-p \left( \frac{\partial g_{\rm poly}}{\partial p} \right)_T. \end{equation}

Differentiating this, we obtain the pressure as a function of specific
volume and temperature: the desired polymorphic equation of state.

\section{Extension of the quasiharmonic method to unstable phonons}

\subsection{Analytic model of soft-mode phonons}

In minerals such as perovskites \cite{warren_phd} it is possible to
describe the transition from a high-temperature phase to a
low-temperature phase of lesser symmetry as the ``freezing in'' of a
finite amplitude of an unstable phonon of the high-symmetry phase, 
plus a finite strain on the unit cell. We
consider the application of quasiharmonic ideas to these materials.

The simple harmonic model gives a negative energy and divergent free
energy arising from the unstable modes. In reality, the soft-mode
phonon is best described by a potential double-well with a local
maximum at the mean structure, corresponding to the high-symmetry
phase.

Let $x_i$ be a coordinate describing the structural feature involved
in the phase transition at a particular wavevector. The corresponding
normal mode can be modeled by considering the dynamics of the set of
$\{x_i\}$ moving in fixed local potential double-wells
\cite{bruce}. In the harmonic limit, normal modes are
uncoupled. However, because we are considering finite displacements
there will in general be coupling between our double-well oscillators,
this being most pronounced around the phase transition and at high
temperatures. Coupling can be approximately treated by renormalization
\cite{bruce}.

Much work has been concentrated on the Landau model in which the
double-well is a {\it free energy\/} in the form of a quartic
polynomial $V(x) = A x^4 - B(T) x^2$, where $B(T)$ changes sign with
temperature through coupling to other modes.

Such a polynomial expansion of the {\it total energy} is also possible,
perhaps incorporating still higher-order terms
\cite{effective}. However, analytic terms beyond second order imply
phonon coupling. This is inconsistent with the harmonic approximation
used to describe non-soft modes: even at high phonon number the normal
modes are assumed to be harmonic and therefore independent of each
other.

We propose instead to describe the entire soft-phonon branch
via a double-well of form:

\begin{equation} V(x) = \frac{1}{2}m\omega_0^2x^2+\epsilon
(e^{-x^2/2\sigma^2}-1). \label{equation_double_well} \end{equation}

where  $\epsilon$, $\omega_0$ and $\sigma$ are wavevector dependent.
Provided that $\epsilon > m \omega_0^2 \sigma^2$, there are minima at:

\begin{equation} x=x_{\pm} \equiv \pm \sqrt{2 \sigma^2 \log (\epsilon /
m\omega_0^2\sigma^2)}, \end{equation}

\noindent separated by a barrier of height:

\begin{eqnarray} \Delta V & \equiv & V(0)-V(x_\pm) \nonumber \\ & =
&\epsilon-m\omega_0^2 \sigma^2 \left( 1 + \log \left( \epsilon /
m\omega_0^2 \sigma^2 \right) \right). \end{eqnarray}

This form of potential has the advantage of being approximately
quadratic in both the low-energy and high-energy limits. Specifically,
for the low-energy case at $x=x_{\pm}$, we have:

\begin{equation} \frac{d^2 V}{dx^2} = 2m \omega_0^2 \log \left( \epsilon/
m \omega_0^2 \sigma^2\right), \end{equation}

\noindent which is equivalent to an harmonic oscillator of frequency:

\begin{equation} \omega_0^\prime =\sqrt{2 \omega_0^2 \log (\epsilon /
m\omega_0^2 \sigma^2)}. \label{equation_low_symm_freq} \end{equation}

\noindent On the other hand, for the high-energy case, the potential
approximates that of an harmonic oscillator of frequency $\omega_0$.

Soft modes do not usually show abnormal dependence on
temperature---except in the vicinity of the phase
transition. Therefore we expect our model to be more widely applicable
than models where soft modes are treated as quartic and other modes as
harmonic.


The (imaginary) harmonic frequency
$\omega_c$ about the (unstable) center of the well is given by:

\begin{equation} \omega_c^2 = \omega_0^2 - \frac{\epsilon}{m \sigma^2}. \label{equation_imaginary_freq} \end{equation}

\subsection{Isolated double-well oscillators}

We now consider the problem of motion in an isolated potential
double-well.

\subsubsection{Classical solution \label{section_classical_solution}}

To evaluate the mechanical energy, we assume the mode is in thermal
contact with a heat bath at the appropriate temperature. The mean
energy is given by:

\begin{eqnarray} \langle E \rangle & = &
\frac{\int_{-\infty}^{\infty}\int_{-\infty}^{\infty} H(p,x) e^{-\beta
H(p,x)} \, dp \, dx}{\int_{-\infty}^{\infty} \int_{-\infty}^{\infty}
e^{-\beta H(p,x)} \, dp\, dx} \nonumber \\ & = & \frac{k_B T}{2} +
\frac{\int_0^\infty V(\sigma z) e^{-\beta V(\sigma z)} \, dz}
{\int_0^\infty e^{-\beta V(\sigma z)} \,dz},
\label{eqn_classical_thermal_energy} \end{eqnarray}

\noindent where $H(p,x)=p^2/2m + V(x)$ is the Hamiltonian of the
isolated mode as a function of $x$ and $p$, the canonical momentum
conjugate to $x$. $\beta = 1/k_B T$ where $k_B$ is Boltzmann's
constant and $T$ is the temperature.

For a given energy $E$, the frequency of our isolated mode can be
evaluated using the action-angle method. The action variable is:

\begin{equation} j \equiv \frac{1}{2\pi} \oint p \, dx. \end{equation}

\noindent If the mode has energy $E \geq \epsilon$, so that there is
sufficient energy to cross the barrier each libration, we find that:

\begin{equation} j = \frac{2\sigma}{\pi}
\int_0^{x_M(E)/\sigma}\sqrt{2mE-m^2 \omega_0^2 \sigma^2 z^2 - 2m
\epsilon e^{-z^2/2}} \, dz,\label{eqn_action_high_energy}
\end{equation}

\noindent where $x_M$ is the positive solution of $V(x)=E$. On the
other hand, if $E < \epsilon$, so that the motion is confined to one
side of the double-well, we find that:

\begin{equation} j = \frac{\sigma}{\pi}
\int_{x_m(E)/\sigma}^{x_M(E)/\sigma} \sqrt{2mE-m^2 \omega_0^2
\sigma^2z^2 - 2m \epsilon e^{-z^2/2}} \, dz,
\label{eqn_action_low_energy}\end{equation}

\noindent where $x_M$ is the greater of the two positive solutions and
$x_m$ is the lesser.

In either case, the corresponding frequency $\omega$ may be evaluated
using Equation \ref{eqn_action_angle}:

\begin{equation} \frac{1}{\omega} = \frac{\partial j}{\partial
E}.\label{eqn_action_angle} \end{equation}

\begin{figure}[h]
\leavevmode \epsfxsize=70mm
\begin{center}
\vspace*{-4.5cm} 
\epsffile{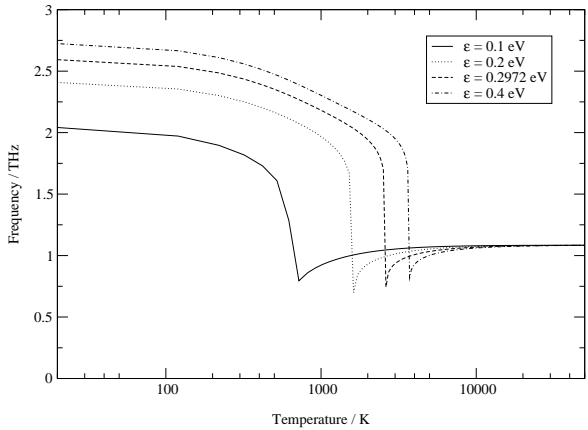}
\end{center}
\caption{Classical frequency of a double-well oscillator in thermal
contact with a heat bath plotted against temperature for various
barrier heights $\epsilon$. The other double-well parameters are:
$m=1$, $\omega_0=0.0691$\,eV$^{1/2}$\,{\AA}$^{-1}$\,amu$^{-1/2}$ and
$\sigma=1.866$\,amu$^{1/2}${\AA}. The frequency falls to zero in a
cusp at the phase transition. At $\epsilon=0.2972$\,eV, the
double-well parameters are appropriate for the double-well describing
the orthorhombic--tetragonal transition in MgSiO$_3$ at zero pressure
\protect\cite{warren_phd}. Thus this model predicts a soft mode transition
temperature of 2609\,K, if the coupling of the soft phonon to strain
is neglected.
\protect\label{figure_thermal_freqs}}
\end{figure}

Taken together, the results of Equations
\ref{eqn_classical_thermal_energy}, \ref{eqn_action_high_energy},
\ref{eqn_action_low_energy} and \ref{eqn_action_angle} allow us to
calculate numerically the frequency of an isolated oscillator moving
with the mean thermal energy as a function of temperature. Example
results are shown in Figure \ref{figure_thermal_freqs}. Typical
``soft-mode'' behavior is observed, with the frequency dropping to
zero in a cusp at the transition temperature. This simple approach was
used in early studies of MgSiO$_3$ \cite{stixrude}.

\subsubsection{Quantum solution \label{section_quantumdw}}

The energy eigenfunctions of a particle moving in a symmetric
potential must be either symmetric or antisymmetric. Furthermore, by
considering building up the Gaussian barrier adiabatically, it is
clear that the symmetry of the eigenfunctions must be the same as for
those of the harmonic oscillator.

The definite symmetry of the wavefunctions leads to a ``paradox'' for
wells of finite separation. If we know the energy of the oscillator
then it is in an energy eigenstate and the wavefunction is either
symmetric or antisymmetric. Hence the probability distribution is
symmetric about the center of the double-well and we cannot
meaningfully say which side the oscillator is confined in, even if its
energy is much less than the barrier height. Thus it is not
conceptually clear that equating the mean thermal energy with the
barrier height gives the correct transition temperature. We discuss
this further in Section \ref{section_dw_phase_tran}.

The Hamiltonian operator for a particle moving in the quadratic
potential {\it without\/} the additional Gaussian potential is:

\begin{equation} \hat{H}^0 = - \frac{\hbar^2}{2m}
\frac{\partial^2}{\partial x^2} + \frac{1}{2}m \omega_0 ^2 x^2 .
\end{equation}

The well-known energy eigenvalues and eigenfunctions for the
time-independent Schr\"{o}dinger equation $\hat{H}^0 \phi_n =
E^0_n\phi_n$ are:

\begin{equation} E^0_n = (n+1/2)\hbar \omega_0 \end{equation}

\noindent and

\begin{equation} \phi_n = \frac{2^{-n/2}}{\sqrt{n!}} \left(
\frac{m\omega_0}{\hbar \pi} \right) ^{1/4} e^{-m \omega_0 x^2 / 2
\hbar} H_n\left( \sqrt{\frac{m \omega_0}{\hbar}} x \right)
\end{equation}

\noindent for $n \in \mathcal{N}_0$, where $H_n(x)$ is the $n$th
Hermite polynomial.

The Hamiltonian operator for a particle moving in the double-well is
$\hat{H}=\hat{H}^0 + V_1$ where $V_1=\epsilon (e^{-x^2/2\sigma^2} -1)$ 
is the extra Gaussian term. Let the eigenfunctions and eigenenergies of
the full Hamiltonian be $\psi_n$ and $E_n$.

The eigenfunctions of the simple harmonic oscillator (SHO) are chosen
as the basis of wavefunction space. This choice makes the computation
particularly simple, as will be seen below.

The matrix elements of the Hamiltonian with respect to our chosen
basis are:

\begin{eqnarray} \langle \phi_i | \hat{H} \phi_j \rangle & = &
\langle\phi_i | \hat{H}^0 \phi_j \rangle + \langle \phi_i | V_1 \phi_j
\rangle\nonumber \\ & = & (i+1/2) \hbar \omega_0 \delta_{i,j}
\nonumber \\ & & +\frac{\epsilon}{2^{(i+j)/2} \sqrt{i!j!\pi}}
\int_{-\infty}^{\infty}e^{-Kz^2} H_i(z) H_j(z) \, dz, \nonumber \\ & &
\end{eqnarray}

\noindent where $K \equiv \hbar /2m\omega_0\sigma^2 + 1$. Note that
the matrix is real and symmetric.

The Hermite polynomials satisfy $H_i(-x)=(-1)^i H_i(x)$, so that
$\langle \phi_i | V_1 \phi_j \rangle = 0$ if $i+j$ is odd. If $i+j$ is
even then the integrand is an even function. Hence, in this case:

\begin{equation} \langle \phi_i | V_1 \phi_j \rangle =
\frac{2\epsilon}{2^{(i+j)/2} \sqrt{i!j!\pi}} \int_0^{\infty} e^{-Kz^2}
H_i(z)H_j(z) \, dz. \label{equation_matrix_of_gaussian} \end{equation}


The eigenvalues of the matrix of the Hamiltonian are the allowed
energy levels. For energies that are large compared with the barrier
height the particle will spend most of its time away from the center
of the well; hence we expect that the system will behave as a SHO in
this limit.  Indeed, it is clear from Equation
\ref{equation_matrix_of_gaussian} that the elements of the matrix
$\langle \phi_i | V_1 \phi_j \rangle$ fall off rapidly as $i$ and $j$
increase. Hence, for large $i$ or $j$, the eigenvalues of the
Hamiltonian tend to to those of the SHO. Thus we only need to
diagonalize the upper left-hand corner (say, the $(n_c+1) \times
(n_c+1)$ submatrix) in order to obtain the first $n_c+1$ energy levels
$E_0$ to $E_{n_c}$. Beyond $n_c$ the eigenvalues may be taken to be
those of the SHO. The comparative ease with which the Hamiltonian
matrix can be diagonalized is one of the advantages of the
quadratic-plus-gaussian double-well over the quartic double-well
potential, although there is no analytic form equivalent to equation
\ref{equation_harmonic_free_energy}

Assuming that $n_c$ is sufficiently large, the canonical partition
function can be written as:

\begin{eqnarray} Z & = & \sum_{n=0}^{n_c} e^{-\beta E_n}
+\sum_{n=n_c+1}^{\infty} e^{-\beta (n+1/2) \hbar \omega_0} \nonumber
\\ &= & \sum_{n=0}^{n_c} e^{-\beta E_n} + \frac{e^{-\beta \hbar
\omega_0(n_c+1)}}{e^{\beta \hbar \omega_0 /2} - e^{-\beta \hbar
\omega_0 /2}}. \label{eq:Zanharm} \end{eqnarray}

The free energy of the double-well oscillator can then be evaluated as:

\begin{equation} F_1=-k_B T \log \left( Z \right).
\label{equation_dw_free_energy} \end{equation}

\subsection{Practical implementation in the quasiharmonic method \label{section_practical_implementation}}


Having proposed that each soft mode at a given wavevector be described
by a double-well of the form given in Equation
\ref{equation_double_well}, we now describe how the parameters
$\epsilon$, $\sigma$ and $\omega_0$ can be determined. Note that if
the $\{ x_i \}$ are mass-reduced phonon coefficients then we may,
without loss of generality, set $m=1$.

Consider the phonon dispersion curve of a crystal structure in which
imaginary frequencies are present. Those branches that remain real
throughout the whole of the Brillouin zone are treated as harmonic
and, for each mode at each wavevector, Equation
\ref{equation_harmonic_free_energy} may be used to find the
corresponding free energy. For those branches that are imaginary in
some region of the Brillouin zone, however, we propose the following
treatment:

\begin{enumerate}

\item  At each symmetry point of the Brillouin zone the eigenvector
corresponding to the relevant mode should be evaluated and the
displacement pattern frozen into the crystal. {\it Ab initio\/}
techniques can then be used to find the corresponding low-symmetry
structure, if desired. Using three total energy calculations with
different amplitudes of the soft phonon frozen into the
structure\footnote{In fact, if we have calculated the total energy of
the high-symmetry phase (corresponding to zero amplitude of the
phonon) then we only need to carry out a further two frozen phonon
calculations.}, we may evaluate the parameters $\epsilon$, $\sigma$
and $\omega_0$ of the double-well. Note that it is possible to fit 
equation \ref{equation_double_well} to every branch even if the
mode is not imaginary
since that Equation
\ref{equation_double_well} does not necessarily describe a
double-well.
In practice the harmonic approximation is 
used for all-real branches, it is equivalent to setting 
$\epsilon=0$.

\item For each branch we use our results for the double-well
parameters at the symmetry points to construct interpolating
polynomials over the whole of the Brillouin zone for the $\epsilon$
and $\sigma$ parameters.

\item For any wavevector in the Brillouin zone, we may find the
spectrum of corresponding (possibly imaginary) frequencies. Provided
we know to which branch these modes belong, we have sufficient
information to determine the parameters of the appropriate double-well
for each mode. $\epsilon$ and $\sigma$ are found by interpolating to
our wavevector and the unstable frequency gives $\omega_c$ (see
Equation \ref{equation_imaginary_freq}), from which we may find
$\omega_0^2=\omega_c^2 + \epsilon / m \sigma^2$.

\item Hence, for any given wavevector, the free energy of each mode,
whether harmonic or soft, can be evaluated. These free energies can be
summed to give the free energy contribution from all modes at the given 
wavevector.

\item 
The free energy can then be integrated over all wavevectors in the Brillouin
zone to give the total lattice thermal free energy. By using a
grid-based scheme to integrate over an irreducible wedge of the zone
and by making use of the continuity of the gradient of each branch, it
is possible to keep track of which branch is which---necessary if the
appropriate values of $\epsilon$ and $\sigma$ are to be
interpolated in the pressence of imaginary branch crossings. 
The problem of interpolation over an irreducible wedge
of the Brillouin zone has been studied extensively in the context of
electronic eigenvalues: see, for example, Reference \cite{blochl}.

\end{enumerate}

The high-symmetry dynamically stabilized phase and the low-symmetry
``frozen-phonon'' phase can now be treated as being distinct. Hence
the methodology of Section \ref{section_polymorphism} can be applied
to find the phase diagram.

\subsection{Interpretation of the soft mode transition
\label{section_dw_phase_tran}}

Consider an isolated symmetric double-well oscillator. For the
probability density to be asymmetric---necessary if we are to
meaningfully say that the particle is in one well or the other---we
must have a mixture of symmetric and antisymmetric energy eigenstates.
Therefore we cannot simultaneously know the energy of our particle
{\it and\/} which well it is in unless we break the symmetry 
(e.g. by allowing the crystal to distort under phonon-strain coupling).

If the oscillating particle's wavefunction is a superposition of
different energy eigenfunctions then the expansion coefficients will
evolve in time (provided the system remains both undisturbed {\it
and\/} unobserved) according to the Schr\"{o}dinger equation. Hence
the quantum mechanical expectation value of the particle's position,
$\langle x \rangle$ changes in time.

The time-dependent wavefunction can be written as:

\begin{equation} \Psi (x,t) = \sum_{n=0}^\infty c_n e^{-i E_nt/ \hbar}
\psi_n (x). \end{equation}

\noindent Hence the expectation of $x$ can be written as:

\begin{eqnarray} \langle x \rangle & = &
\sum_{n=0}^\infty\sum_{m=0}^\infty c_n^\ast c_m e^{i (E_n-E_m)t/\hbar}
\langle \psi_n | x\psi_m \rangle \nonumber\\ & = & 2 \sum_{n \, {\rm
even}} \sum_{m \, {\rm odd}} \left| c_n \right| \left| c_m \right|
\nonumber\\ & & \hspace{1.8cm} \times \cos (\omega_{nm} t +\eta_{nm})
\langle \psi_n | x \psi_m \rangle \end{eqnarray}

\noindent where $\omega_{nm}=|E_n-E_m|/\hbar$ and
$\eta_{nm}=(\arg(c_m)-\arg(c_n)) {\rm sgn}(\omega_{nm})$. Note that we
use the fact that $\langle \psi_n | x \psi_m \rangle =0$ if $\psi_n$
and $\psi_m$ are either both odd or both even since $x$ itself is
odd. We also use the fact that $\langle \psi_n | x \psi_m
\rangle=\langle \psi_m | x \psi_n \rangle$.

We assume that the energy levels are initially populated according to
Boltzmann statistics; thus $|c_n|^2=Z^{-1}e^{- \beta E_n}$, where
$\beta = 1/k_B T$ and $Z=\sum_{n=0}^\infty e^{- \beta E_n}$ is the
canonical partition function.

So we have:

\begin{equation} \langle x \rangle = \sum_{n \, {\rm even}} \sum_{m \,
{\rm odd}} \Gamma_{nm} \cos (\omega_{nm} t +\eta_{nm}), \end{equation}

\noindent where $\Gamma_{nm} = 2 Z^{-1} e^{-\beta (E_n+E_m)/2}
\langle\psi_n | x \psi_m \rangle$ is the amplitude of the sinusoidal
component in the expansion of $\langle x \rangle$ with frequency
$\omega_{nm}$.

For the harmonic oscillator potential, the frequencies of the
oscillations in $\langle x \rangle$ are of the form
$\omega_{nm}=|E_n-E_m|/\hbar = |n-m| \omega_0$. Thus the lowest
oscillation frequency is $\omega_0$. For the symmetric double-well,
however, we end up with a set of pairs of energy levels that are very
close to each other (becoming degenerate in the limit that the barrier
height goes to infinity). These give rise to oscillation frequencies
very much lower than $\omega_0$.

As the temperature is increased, higher frequency components have
larger $\Gamma$ coefficients. We suggest that the soft mode phase
transition be judged to occur when the frequency with the highest
coefficient exceeds the frequency of the experimental probe. When this
has happened, the predominant sinusoidal component of $\langle
x\rangle$ has a frequency higher than can be measured by the
experimental probe, and so it appears to the experimenter that
$\langle x \rangle = 0$. Below this temperature, measurements of
$\langle x \rangle$ will tend to find it in one well or the
other. This definition is different from the polymorphic one (Section
\ref{section_polymorphism}) because of the contribution to the free
energy from the symmetry-breaking distortion of the lattice that
inevitably accompanies the transition. In particular, the
quasiharmonic transition is first-order while this measurement dependent 
``soft mode'' one is second-order \cite{bruce}. 

Thus the first-order transition is determined by comparing:

\begin{enumerate}

\item  The free energy of the soft-mode phase, calculated as above,  
expanded about an unstable frozen-ion structure
without strain-phonon coupling.

with 

\item The free energy of the low symmetry phase, calculated by 
expanding about the minimum of total energy.

\end{enumerate}

Typically, the former will have higher entropy (sampling from both wells)
while the latter has lower energy.

\subsection{Absorption of low-frequency photons}

\begin{figure}
\begin{center}
 \epsfxsize=70mm
\epsffile{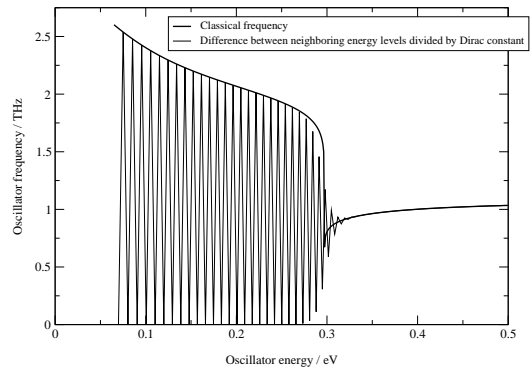}
\end{center}
\caption{Thick line: Frequency of classical isolated double-well oscillator
against energy. The double-well parameters are as for Figure
\ref{figure_thermal_freqs} with $\epsilon=0.2972$\,eV. Fine line: Energy
differences $(E_i -E_{i-1})/\hbar$ against energy $(E_i+E_{i-1})/2\hbar$
for quantum oscillator. \label{figure_dw_freq_v_energy}}
\end{figure}

Figure \ref{figure_dw_freq_v_energy} shows the energy difference
between neighboring energy levels $(E_n-E_{n-1})/\hbar$ plotted
against the mean of the two energies $(E_n+E_{n-1})/2$ for the quantum
double-well oscillator. Absorption of photons at frequency
$(E_n-E_{n-1})/\hbar$ is symmetry-allowed.

For energies in excess of the barrier height the frequency
$(E_n-E_{n-1})/\hbar$ is virtually identical to the classical
frequency for energy $(E_n+E_{n-1})/2$, obtained using the method of
Section \ref{section_classical_solution}. In the very high-energy
limit the frequency behaves as that of the quadratic potential well
without the Gaussian barrier.

For energies less than the barrier height the energy levels tend to
degenerate pairs of levels. The frequencies given by the difference
between the energy levels of neighboring pairs again correspond to the
classical frequencies. However, the pairs of almost-degenerate
eigenstates imply the existence of very low-frequency absorption
peaks. (These are the very low frequencies that alternate with the
classical frequencies below the transition energy in Figure
\ref{figure_dw_freq_v_energy}.) It should be noted that these
frequencies are not associated with the normal modes of the
low-symmetry phase and do not, therefore, contribute to the
quasiharmonic thermal energy. They are a feature of the quantum
double-well oscillator without classical analog.

\section{Application of quasiharmonic methods to periclase}

\subsection{Computational details}

\subsubsection{The cold-curve}

For each lattice parameter the total energy is evaluated using the
CASTEP software package \cite{payne_castep}, which utilizes
density-functional theory in the generalized gradient approximation
(GGA) \cite{perdew}. The ionic cores are accounted for using ultrasoft
pseudopotentials \cite{vanderbilt} in Kleinman-Bylander form
\cite{payne_rmp}. The wavefunctions of the valence electrons are
expanded in a plane wave basis set up to an energy cutoff of 540\,eV.

For the B1 phase, the simulation cell consists of a single cubic unit
cell. The Brillouin zone is sampled at 20 special points generated
from an $8 \times 8 \times 8$ mesh using the Monkhorst-Pack scheme
\cite{monkhorst}. For the B2 phase, the simulation cell is a single
cubic primitive cell. The Brillouin zone is sampled at 35 special
points from a $9 \times 9 \times 9$ mesh. In each case the point
symmetries of the crystal are enforced \cite{kunc}.


The equilibrium lattice parameter of the B1 phase at zero external
pressure (which corresponds to the minimum of the cold-curve)
calculated using CASTEP is $a=4.259$\,{\AA}, which may be compared
with an experimentally determined parameter $a=4.2115(1)$\,{\AA}
\cite{reichmann}. The difference between the theoretical and
experimental values is about 1\%.

\subsubsection{Determination of the matrix of force constants}

The supercells simulated to determine the matrices of force constants
for the B1 phase consist of $2 \times 2 \times 2$ cubic unit cells (64
atoms). Thus the interactions between a given ion and its
third-closest shell of neighbors are included in our calculations. For
the B2 phase, the cells used consist of $2 \times 2 \times 2$ cubic
primitive unit cells (16 atoms). In these supercells the crystal
symmetry is such that only two ionic displacements are required to
complete the entire matrix of force constants. The plane-wave cutoff
energy is 540\,eV and the Brillouin zone is sampled at 6 special
points from a $4 \times 4 \times 4$ mesh. In each simulation the ion
displaced from equilibrium is moved by $0.4\%$ of the lattice
parameter.

As demonstrated by Parlinski \cite{parlinski_mgo} it is possible to
calculate the Born effective charge tensors from first principles
using simulations of elongated supercells.  Note that because of the
symmetry of the B1 and B2 phases, the Born effective charge tensors
are isotropic (so $Z_{n,\alpha,\beta}=Z_n
\delta_{\alpha,\beta}$). Furthermore, the sum of the Born effective
charges over the ions in a unit cell must be zero \cite{cochran} (so
$Z_{\rm Mg}=-Z_{\rm O}$). Hence there is effectively only one
undetermined parameter in the non-analytic term: $Z_{\rm Mg}({\bf 0})
/ \sqrt{\epsilon_0({\bf 0})}$.

\begin{figure}[h]
\leavevmode \epsfxsize=70mm
\begin{center}
\vspace*{-4.0cm} 
\epsffile{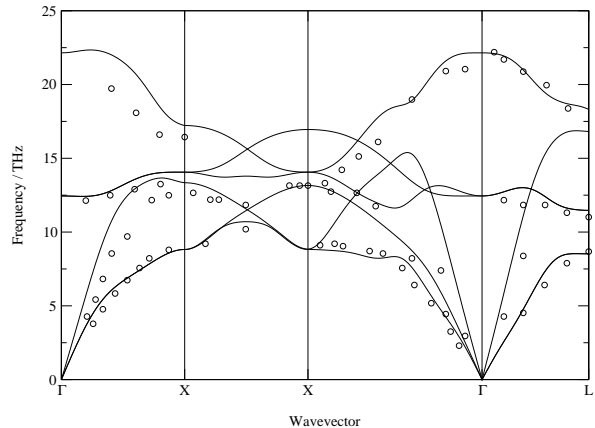}
\end{center}
\caption{Dispersion curves for B1 MgO at lattice parameter 4.2\,{\AA}.
The symmetry points in the dispersion curve are (from left to right):
$\Gamma$ [000], X [001], X [011], $\Gamma$ [000] and L $[\frac{1}{2}
\frac{1}{2} \frac{1}{2}]$. Note the LO--TO splitting (At $\Gamma$, LO
frequency is 22.0\,THz whereas TO frequency is 12.4\,THz) arising from
the non-analytic term of Equation \ref{eq:non_analytic}. Also shown
are Peckham's experimental results for B1 MgO at lattice parameter
4.212\,{\AA}. Note that the reciprocal lattice vectors referred to in
the results for the B1 phase are those of the cubic unit cell rather
than the true reciprocal lattice vectors.
\label{fig:mgo_B1_dispersion}}
\end{figure}

We choose $Z_{\rm Mg}({\bf 0}) / \sqrt{\epsilon_0({\bf 0})}=4.4$ to
give the LO branch in the dispersion curve of Figure
\ref{fig:mgo_B1_dispersion}. Our values for the effective charge
tensors of the B1 phase are such that the calculated LO branch for
lattice parameter 4.2\,{\AA} are in reasonable agreement the
experimental results of Peckham \cite{peckham}. We neglect the
variation of the effective charge tensors with lattice parameter.

\subsection{Approximations and errors}

\subsubsection{Errors in \textit{ab initio} total energy calculations}

Total energy differences between structures calculated using
density-functional theory in the generalized gradient approximation
are thought to be reliable to within a few percent \cite{payne_rmp}.

The cutoff energy of the plane-wave basis at 540\,eV is sufficient for
convergence of the total energies of the crystals to within
10\,$\mu$eV per ion, several orders of magnitude less than the likely
error due to the use of the GGA. The Hellmann-Feynman forces are
converged to within 1\,meV\,{\AA}$^{-1}$, at least two orders of
magnitude less than the dominant forces arising when an ion is
displaced.

\subsubsection{The harmonic approximation}

We investigate the range of validity of the harmonic
approximation. This is done for the B1 phase with lattice parameter
$a=4.2$\,{\AA}.

\begin{figure}[h]
\leavevmode \epsfxsize=70mm
\begin{center}
\vspace*{-3.5cm} 
\epsffile{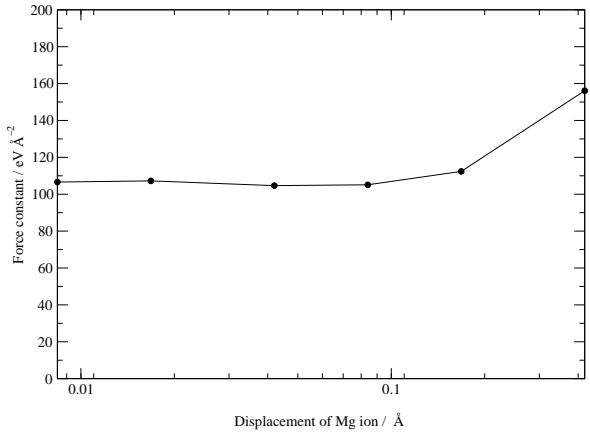}
\end{center}
\caption{Graph of restoring force divided by the finite displacement
of a Mg ion in the [100]-direction against that displacement. Results
are appropriate for a $2 \times 1 \times 1$ supercell of B1 phase at
lattice parameter 4.2\,{\AA}. The force constant is independent of the
magnitude of the displacement until the displacement is at least
0.084\,{\AA}.  Thereafter the restoring force increases faster than
linearly with the displacement; the harmonic approximation has broken
down.
\label{fig:test_harm}}
\end{figure}

We evaluate the force constant of the restoring force on a magnesium
ion as it is displaced in the $x$-direction. The results are shown in
Figure \ref{fig:test_harm}. It can be seen that the force constant
starts to increase when the ionic displacement reaches $a_{\rm max}
\approx 0.084$\,{\AA}, about 2\% of the lattice parameter, at which
point the potential energy is about 0.4\,eV. Other displacements are
similar; hence it is reasonable to assume that the forces remain
linear (and the quasiharmonic assumption is valid) for temperatures up
to several thousand Kelvin.

\subsubsection{Other sources of error}

Another potential source of error is the limited size of the
simulation supercells. However, the results shown in Table
\ref{table_range_falloff} for the B1 phase make it clear that
interactions beyond the third-closest shell of neighbors may be safely
neglected \cite{nonanal}.

Contributions to the free energy from the thermal excitation of the
valence electrons, from coupled electron-phonon excitations and from
the equilibrium population of defects are thought to be negligible in
comparison with the frozen-ion and lattice thermal energies
\cite{swift_phd}.

Poor convergence of the Hellmann-Feynman forces can result in the
violation of Newton's third law for the matrix of force constants.
Typically this results in the acoustic branches of the dispersion
curve failing to pass through zero at the center of the Brillouin zone
\cite{ackland}. As discussed in Section \ref{section_latt_thermal},
Newton's third law is imposed on the matrix of force
constants. However, even without this, the calculated acoustic
branches pass very close to zero at the zone center.

The method by which long-range polarization effects are accounted for
is also approximate. The effects of this are discussed below.

\subsection{\textit{Ab initio} phonons}

\subsubsection{The B1 phase}

Inelastic neutron scattering experiments were carried out by Peckham
\cite{peckham} and used to generate dispersion curves for the B1 phase
\cite{sangster}. We compare our theoretical dispersion curve with
these results in Figure \ref{fig:mgo_B1_dispersion}. (Note that our
dispersion curve was generated for lattice parameter 4.2\,{\AA},
whereas Peckham's results were obtained under ambient conditions where
the lattice parameter is 4.212\,{\AA}.) Our theoretical results are in
reasonable agreement with experiment. (The lattice parameter of
$4.2$\,{\AA} corresponds to a pressure of about 7\,GPa at zero
temperature.)

\begin{figure}[h]
\leavevmode \epsfxsize=70mm
\begin{center}
\vspace*{-3.5cm} 
\epsffile{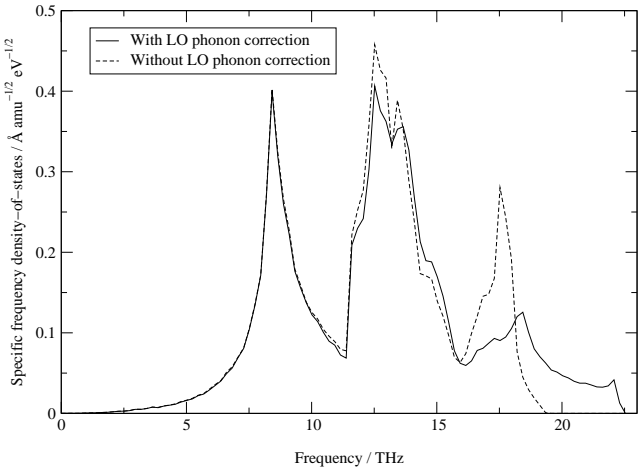}
\end{center}
\caption{Specific frequency density-of-states for B1 phase at lattice
parameter 4.2\,{\AA} with and without the inclusion of the
non-analytic term in the dynamical matrix. \label{fig:mgo_B1_freq_DoS}}
\end{figure}

The specific frequency density-of-states function is shown in Figure
\ref{fig:mgo_B1_freq_DoS}. Without the addition of the non-analytic
term to the dynamical matrix, the longitudinal optic branch is
degenerate with the transverse optic branch at the $\Gamma$-point, and
this is also shown.  Although only the LO branch is altered
substantially, it can be seen that the inclusion of the non-analytic
term has a significant effect on the density of states.

We compare sound velocities calculated from our dispersion curves with
the experimental results of Reichmann {\it et al} \cite{reichmann}
obtained using ultrasonic interferometry. Reichmann obtains a P-wave
sound speed of 9119\,ms$^{-1}$ in the $[100]$-direction whereas our
longitudinal-acoustic mode has $\left( d\omega_{\rm LA} / dk_{[100]}
\right)_{{\bf k}=0}=11367$\,ms$^{-1}$.

On the other hand, in the $[111]$-direction, the experimental P-wave
velocity is 10125\,ms$^{-1}$ which may be compared with our
theoretical value of 10818\,ms$^{-1}$.

Thus Reichmann's experimental results show a higher degree of
anisotropy than do our theoretical results. The discrepancy would
appear to be (at least partly) caused by the imposition of symmetry
and Newton's third law on the matrix of force constants
\cite{ackland}: if this procedure is not carried out then we find that
$\left( d\omega_{\rm LA} / dk_{[100]} \right)_{{\bf
k}=0}=10770$\,ms$^{-1}$ and $\left( d\omega_{\rm LA} / dk_{[111]}
\right)_{{\bf k}=0}=12651$\,ms$^{-1}$. Although Reichmann's P-wave
velocities are still somewhat less than these theoretical velocities,
both are now in agreement that the velocity in the $[100]$-direction
is less than the velocity in the $[111]$-direction.

\subsubsection{The B2 phase}

\begin{figure}[h]
\leavevmode \epsfxsize=70mm
\begin{center}
\vspace*{-3.5cm} 
\epsffile{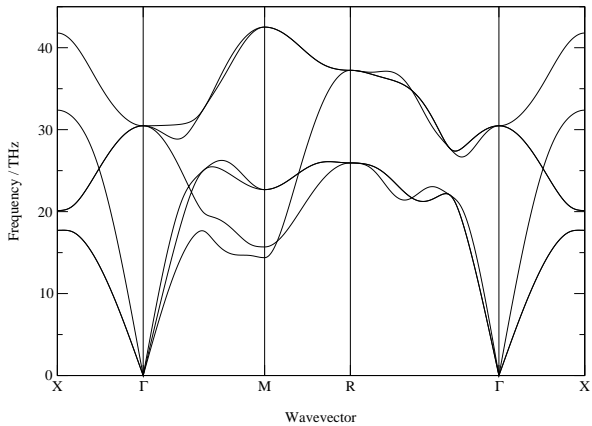}
\end{center}
\caption{Dispersion curves for B2 MgO at a lattice parameter
2.0\,{\AA}.  The symmetry points in the dispersion curve are (from
left to right): X $[\frac{1}{2} 00]$, $\Gamma$ [000], M $[\frac{1}{2}
\frac{1}{2} 0]$, R $[\frac{1}{2} \frac{1}{2} \frac{1}{2}]$, $\Gamma$
[000] and X $[\frac{1}{2} 00]$. \label{fig:mgo_B2_dispersion}}
\end{figure}
\begin{figure}[h]
\leavevmode \epsfxsize=70mm
\begin{center}
\vspace*{-3.5cm} 
\epsffile{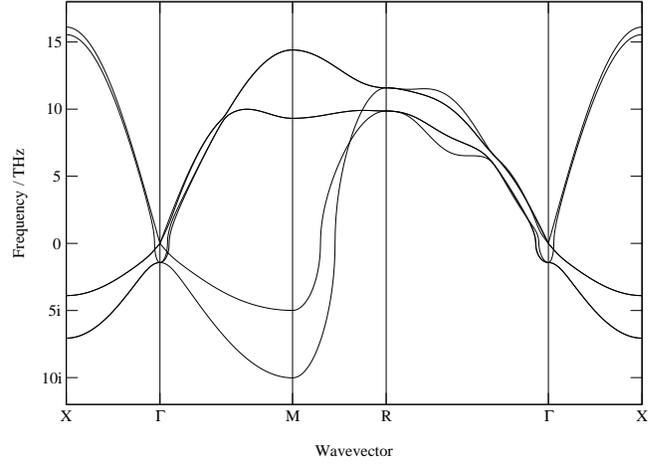}
\end{center}
\caption{Dispersion curves for B2 MgO at a lattice constant of
2.7\,{\AA}. The symmetry points are as for Figure
\ref{fig:mgo_B2_dispersion}. Note the branches of imaginary phonon
frequencies, indicating that the structure is mechanically unstable at
this volume. \label{fig:mgo_B2_imaginary}}
\end{figure}

Typical dispersion curves for the B2 phase at lattice parameters 2.0\,{\AA}
and 2.7\,{\AA} are shown in Figures \ref{fig:mgo_B2_dispersion} and
\ref{fig:mgo_B2_imaginary}. At zero temperature these lattice
parameters correspond to pressures of $653$\,GPa and $-6$\,GPa
respectively.


Note the presence of unstable modes at low pressures in the dispersion
curve of the B2 phase (Figure \ref{fig:mgo_B2_imaginary}). We find
that the B2 phase is structurally unstable for pressures below about
82\,GPa.  In our
calculations for the phase boundary in periclase we do {\it not\/} 
require our novel method for
dealing with soft modes because the imaginary frequencies are only
found in the B2 phase at pressures for which the B1 phase is clearly
favored.

\subsection{\textit{Ab initio} equation of state}

\begin{figure}[h]
\leavevmode \epsfxsize=70mm
\begin{center}
\vspace*{-3.5cm} 
\epsffile{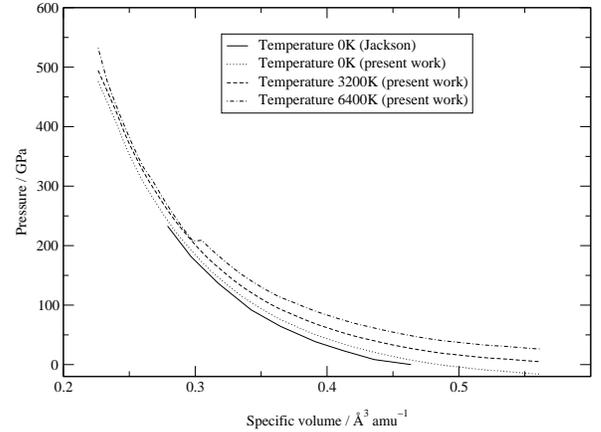}
\end{center}
\caption{Equation of state of periclase: pressure against specific
volume at various temperatures. The position of the phase transition
is clearly visible as a kink in the curves. A Birch-Murnaghan equation
of state generated from Jackson's experimental results is also shown.
\label{fig:mgo_pressures}}
\end{figure}

We plot the pressure against specific volume for a range of
temperatures in Figure \ref{fig:mgo_pressures}. This is the desired
thermodynamic equation of state for periclase. Also shown is a
third-order Birch-Murnaghan equation of state generated from the
isothermal bulk modulus and its first derivative with respect to
volume, which were obtained by means of ultrasonic sound velocity
measurement \cite{jackson}.

As a test of the validity of our results, the isothermal bulk modulus
at zero temperature and external pressure (where periclase is entirely
in the B1 phase) can be compared with experimental results. We
calculate the bulk modulus to be $-v \left( \partial p / \partial v
\right)_{T,v} = 155$\,GPa, whereas an experimentally determined
value is $160.3 \pm 0.3$\,GPa \cite{jackson}. The theoretical and
experimental results differ by about 3\%.


We may also compare the pressure derivative of the bulk modulus at
zero pressure and temperature with experimental results. The bulk
modulus is found to be almost, but not quite, a linear function of
pressure. Fitting a straight line to the data from $-8.3$\,GPa to
$21.12$\,GPa we find that the the gradient is $4.11$, in excellent
agreement with the experimentally determined value of $4.2 \pm 0.2$
\cite{jackson}. It is found that the pressure derivative of the bulk
modulus decreases slightly as the pressure is increased.

\subsection{\textit{Ab initio} phase diagram}

\begin{figure}[h]
\leavevmode \epsfxsize=70mm
\begin{center}
\vspace*{-3.5cm} 
\epsffile{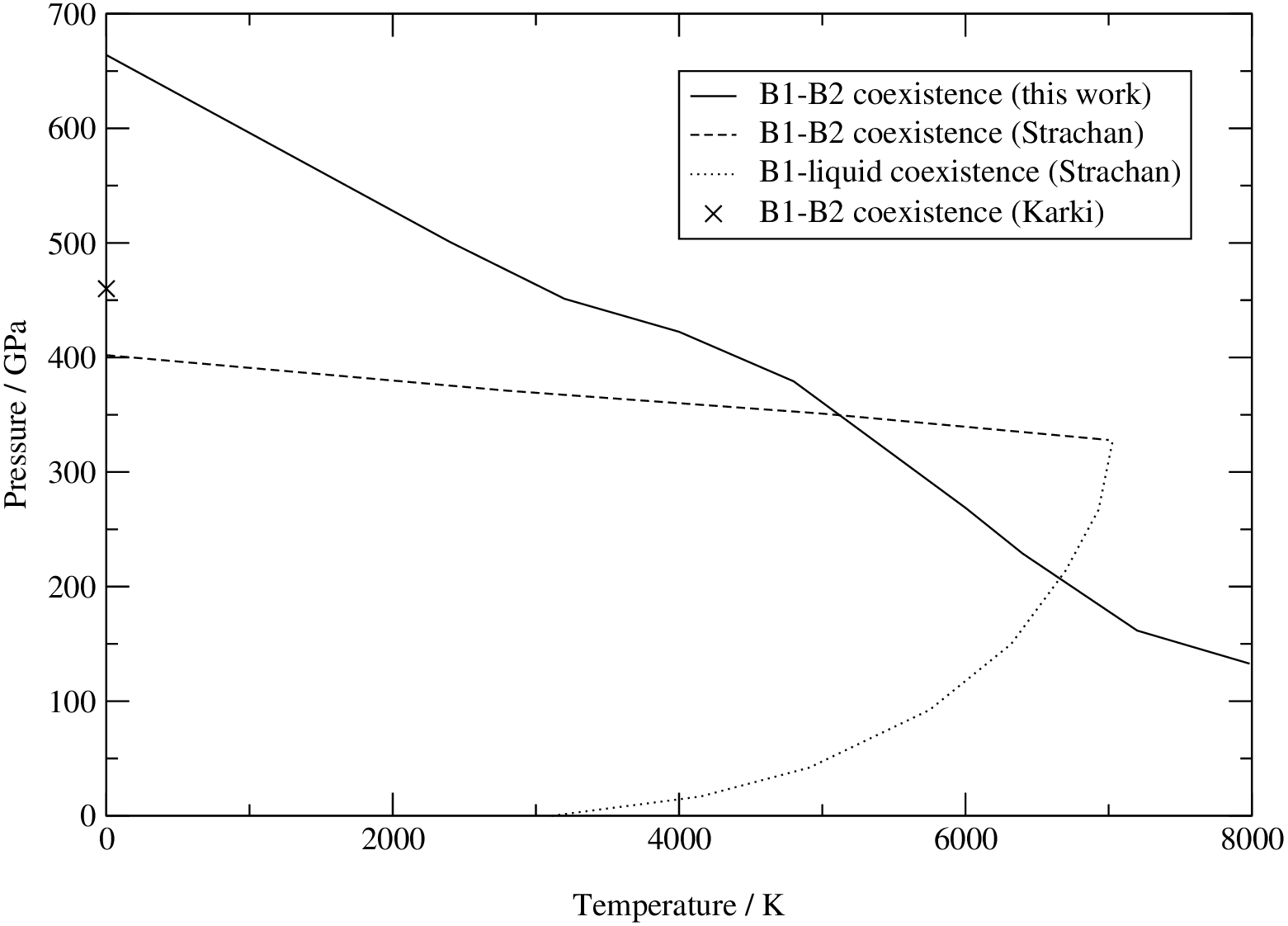}
\end{center}
\caption{Theoretical phase diagram of periclase. The B1 phase region
is below and to the left of the B1--B2 coexistence line. Also shown
are other theoretical results\protect\cite{karki,strachan}, 
including a theoretical B1--liquid
phase boundary. \protect\label{fig:phase_diag}
}
\end{figure}

The theoretical phase diagram of solid periclase is shown in Figure
\ref{fig:phase_diag}. At pressures and temperatures below and to the
left of the phase boundary shown in the diagram, periclase exists in
the B1 phase; above and to the right of the boundary it exists in the
B2 phase.  Duffy \cite{duffy} has shown experimentally that, at room
temperature, the B1 phase is stable to pressures of at least
$227$\,GPa. This lies well within the B1 region of our theoretical
phase diagram. It can also be noted that the B2 phase is not favored
at the pressures at which it is structurally unstable.

\begin{figure}
\vspace{-0.5cm}
 \epsfxsize=70mm
\epsffile{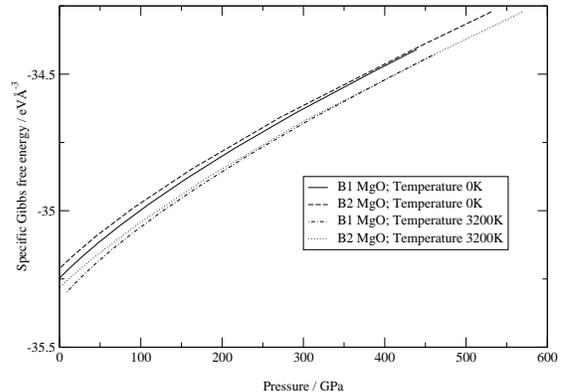}
\caption{Gibbs free energy plotted against pressure for the B1 and B2
phases of periclase at two different temperatures. The transition
pressure at each temperature corresponds to the point where the curves
cross. The similarity of the curves for each phase means that small
errors in the energies lead to large uncertainties in the transition
pressures.
\label{fig:gibbs_crossing}}
\end{figure}

Also shown in Figure \ref{fig:phase_diag} is the theoretical phase
diagram obtained by Strachan {\it et al\/} \cite{strachan} using
molecular dynamics simulation. Although qualitatively similar, the
calculated position and orientation of the B1--B2 phase boundary is
very different. Karki \cite{karki} obtained a B1--B2 transition
pressure of 460\,GPa at 0\,K, which also differs substantially from
that of Strachan. Figure \ref{fig:gibbs_crossing} shows the Gibbs free
energy plotted against pressure for the two phases at two different
temperatures. The difficulty in ascertaining the transition pressures
at which the curves cross is apparent. This consideration will affect
all theoretical calculations of the phase diagram of periclase.

\begin{figure}
 \epsfxsize=70mm
\epsffile{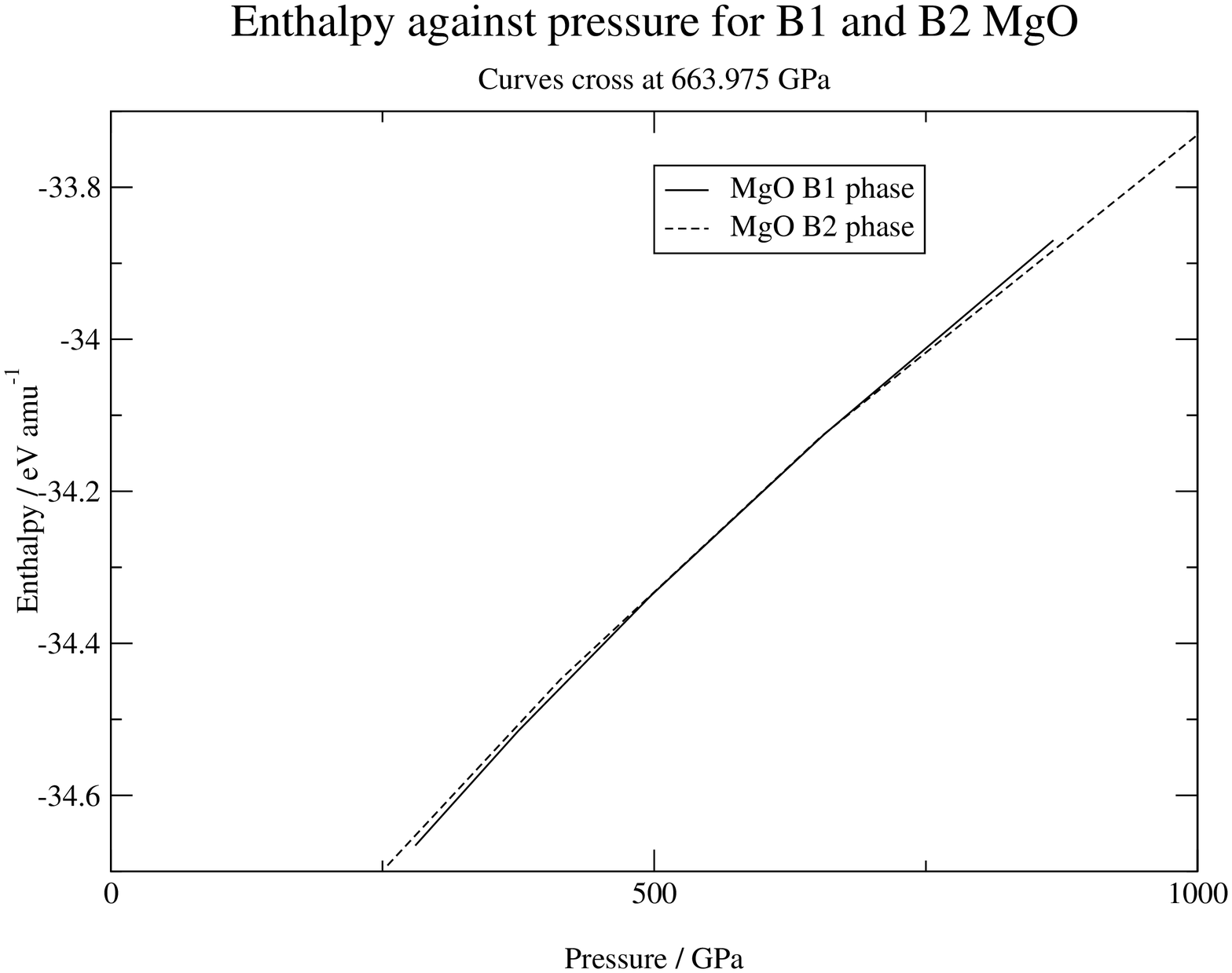}
\caption{Enthalpy $h=e+pv$ plotted against pressure for the B1 and B2
phases of MgO at temperature 0\,K (note that zero-point energy is not
included in the results shown in this graph). The curves cross (and so
the phase transition is predicted to occur) at 664\,GPa. These
results are obtained directly from the results of CASTEP. Note that,
again, the two curves are extremely close when they cross; thus the
position of the theoretical transition point is sensitive to the
simulation parameters.
\label{fig:mgo_enthalpies}}
\end{figure}

We confirm the difficulty in locating the transition by attempting to
reproduce Karki's zero-temperature results in which zero-point lattice
vibrational energy is neglected. We simply calculate the enthalpy
against pressure for the two phases using CASTEP. The results are
shown in Figure \ref{fig:mgo_enthalpies}. We find a transition
pressure of \,GPa, different from that of Karki (460\,GPa). The
possibility of a metallization transition lowering the energy of the
B2 phase was investigated but found not to occur at relevant pressures.

We consider the fractional error in unit cell volume for the B2 phase
to which the difference between the specific enthalpies of the B1 and
B2 phases at 451\,GPa corresponds. For given unit cell enthalpy for
the B2 phase, we find the fractional volume difference as:

\begin{equation} \left| \frac{\Delta V}{V} \right| = 2 \left| \frac{h_{B2}-h_{B1}}{h_{B2}+h_{B1}} \right|, \end{equation}

\noindent where $h_{B1}=-34.4043$\,eV{\AA}$^{-3}$ and
$h_{B2}=-34.3978$\,eV{\AA}$^{-3}$ are the specific enthalpies of the
two phases at 451\,GPa. Thus a discrepancy of $\Delta V / V 
\times 10^{-4} \approx 0.02$\% in the volume of either phase would
correspond to a 200GPa change in the transition pressure. This
illustrates the sensitivity of the transition point to the details of
how the total energy calculations are carried out.

We use the generalized gradient approximation
 and ultrasoft pseudopotentials\cite{perdew,vanderbilt},  
whereas Karki used the
local density approximation and $Q_c$ tuned pseudopotentials
\cite{lda,qc}. 
The difference between calculated volume 
using these two methods is typically of the order of 5\%; hence
this is likely to be responsible for the difference between our
results and Karki's.



The Clausius-Clapeyron equation for the coexistence line between two
phases in $(p,T)$-space is:

\begin{equation} \frac{dp}{dT} = \frac{\Delta s}{\Delta v}, \end{equation}

\noindent where $p(T)$ is the coexistence line and $\Delta s$ and
$\Delta v$ are the specific entropy and volume differences between the
two phases across the line. At zero temperature the entropy of the two
phases should be zero by the third law of thermodynamics which is valid within our method (though not, e.g. in classical MD\cite{strachan}); therefore,
provided the phases have different densities, the coexistence line
should satisfy $dp/dT=0$ at $T=0$. Figure \ref{fig:phase_diag}.) does 
not appear to satisfy this requirement.

The explanation for this apparent discrepancy lies with the fact
that the densities of the two phases are very similar (and converging) 
at their
predicted zero-temperature transition point: the specific volumes of
the B1 and B2 phases are 0.202\,{\AA}$^3$amu$^{-1}$ and
0.198\,{\AA}$^3$amu$^{-1}$ respectively.

\section{Conclusions}

We have described an extension to the quasiharmonic method that allows
the free energy contribution from ``soft'' phonons in dynamically
stabilized crystals to be evaluated. Our approach is based on a form
of potential double-well different to that used in previous work on
soft phonons: a parabola-plus-Gaussian form that has the advantage of
being harmonic in both the low- and high-temperature limits.

We argue that the first-order nature of the phase transitions found
using our extended quasiharmonic method arises because of the coupling
of the relevant phonon to strain in the crystal. Without this
coupling, the transition would be second-order. We have suggested a
criterion for judging when a second-order soft-mode phase transition
has occurred, taking into account the quantum mechanical nature of the
problem.

At energies less than the height of the central barrier in our
symmetric potential double-well, the allowed energy levels consist of
near-degenerate pairs. Hence we suggest there must exist extremely
low-frequency photon absorption peaks for soft-mode materials in their
low-temperature phase, corresponding to photon-induced transitions
between such pairs.

We have evaluated the equation of state and the phase boundary for the
B1--B2 transition in periclase using {\it ab initio\/} calculations in
the quasiharmonic approximation. We predict that this transition will
occur, but that it is well outside the ranges encountered inside the
Earth. Locating the B1--B2 phase boundary with precision is difficult,
however, because the Gibbs free energy curves of the two phases are
very similar when they cross.

\section{Acknowledgments}

We would like to thank D.~C.~Swift and M.~C.~Warren for helpful
discussion. This work was carried out as part of the UKCP-MSI
collaboration, supported by EPRSC GR/N02337. Copies of the codes used
for the free energy calculations are available from the authors on
request.

%
%

%
%

\begin{table}[h]
\begin{tabular}{|l|l|}
Position of Mg atom & Magnitude of [001]-component \\ & of force /
 eV{\AA}$^{-1}$ \\ \hline (0,0,0) & 0.13424 \\ (0.5,0.5,0.0625) &
 0.03356 \\ (0,0,0.125) & 0.00981 \\ (0.5,0.5,0.1875) & 0.00006
\end{tabular}
\caption{Magnitude of the component of force in the [001]-direction
when the Mg ion at (0,0,0) is displaced in the [001]-direction by
0.05\% of the length of a $1/\sqrt{2} \times 1/\sqrt{2} \times 8$
supercell.  Coordinates are given as fractions of the supercell
dimensions. (These results are for B1 MgO at lattice parameter
4.2\,{\AA}.)
\label{table_range_falloff}}
\end{table}

\end{multicols}


\begin{references}
\bibitem{nonanal} Apart from the non-analytic dipole term, which is
calculated separately.
\end{references}
\end{document}